\documentclass[final,5p,times,twocolumn,authoryear]{elsarticle}

\usepackage{amssymb}
\usepackage{aas_macros}
\usepackage{epsfig}
\usepackage{rotating}

\usepackage{subfig}
\usepackage{amsmath}
\usepackage{float}
\usepackage{multirow}
\usepackage{array}
\usepackage{graphicx}
\usepackage{float}
\usepackage{rotating}
\usepackage{blindtext}
\usepackage[english]{babel}
\usepackage[flushleft]{threeparttable}
\usepackage{tabularx,graphicx}
\usepackage{siunitx}
\usepackage{color}
\usepackage{orcidlink}
\usepackage{listings}
\usepackage{xcolor}
\usepackage{hyperref}
\newcommand{\orcid}[1]{\orcidlink{#1}{\textcolor[HTML]{A6CE39}{}}}
\hypersetup{
    colorlinks=true,
    linkcolor=purple, % Color de los enlaces internos
    citecolor=purple, % Color de las citas
    urlcolor=purple   % Color de los enlaces externos
}

\newcommand{\eagle}[1]{\textit{EAGLE}}
\newcommand{\flamingo}[1]{\textit{FLAMINGO}}
\usepackage{comment}

\journal{New Astronomy}
\usepackage[switch]{lineno}
\renewcommand{\linenumberfont}{\normalfont\tiny\color{magenta}}
\makeatletter
\def\makeLineNumberLeft{%
  \linenumberfont\llap{\hb@xt@\linenumberwidth{\LineNumber\hss}\hskip\linenumbersep}% left line number
  \hskip\columnwidth% skip over column of text
  \rlap{\hskip\linenumbersep\hb@xt@\linenumberwidth{\hss\LineNumber}}\hss}% right line number
  
\definecolor{violet}{rgb}{0.54, 0.17, 0.89}
\definecolor{darkgreen}{rgb}{0.0, 0.5, 0.0} 

\definecolor{darkcyan}{rgb}{0.0, 0.55, 0.55}

\usepackage{blindtext}
\usepackage[normalem]{ulem}

\begin{document}
\def\pdfname{article}

\begin{frontmatter}

\title{Exploring the halo occupation distribution for moderate X-ray luminosity\\ active galactic nuclei in the \eagle{} cosmological simulation}

\author[ia,uabc]{Liliana Altamirano-D\'evora\corref{cor}\orcid{0000-0001-7715-2182}}
\ead{lili@astro.unam.mx}
\author[ia]{Hector Aceves\orcid{0000-0002-7348-8815}}
\author[ia]{Angel Castro\orcid{0000-0002-7832-5337}}
\author[ia]{Takamitsu Miyaji\orcid{0000-0002-7562-485X}}

\cortext[cor]{Corresponding author.}
\address[ia]{Instituto de Astronom\'ia, Universidad Nacional Aut\'onoma de M\'exico, Carretera Tijuana-Ensenada Km 107, Ensenada, 22860, M\'exico}
\address[uabc]{Facultad de Ingenier\'ia, Arquitectura y Dise\~no (FIAD), Universidad Aut\'onoma de Baja California, Carretera Tijuana-Ensenada Km 103, Ensenada, 22860, M\'exico}

\begin{abstract}
The hydrodynamical cosmological simulation \eagle{} is used to model the Halo Occupation Distribution (HOD) of moderate X-ray luminosity active galactic nuclei (mXAGN), extending previous work using only dark matter simulations and empirical relations. By examining mergers as a triggering mechanism, we focus on halos typical of galaxy groups and cluster-like systems with masses $\geq 10^{12.75}\,{\rm M}{\odot}\,h^{-1}$. We analyze simulation data to create catalogs of central and satellite galaxies. We study their merger history we quantify the percentage of minor and major mergers in the mXAGN sample. We obtain the HOD for central and satellite mXAGN across a redshift interval from \(z=2\) to the present epoch. Our results indicate that, across most redshifts, minor mergers slightly predominate as the primary mechanism for triggering mXAGN.
\end{abstract}


\end{frontmatter}

%%%%%%%%%%%%%%%%%%%%%%%%
\section{Introduction} 
%%%%%%%%%%%%%%%%%%%%%%%%
Active Galactic Nuclei (AGN) are astronomical systems where multiple physical processes occur simultaneously, making their theoretical and observational study challenging and highly dynamic. They exhibit extreme luminosity at their centers due to material accreting onto a black hole, accompanied by strong emission lines. These features distinguish them from normal galaxies \citep[e.g.,][]{Seyfert+43, Peterson+97}. Research on AGN is crucial for advancing our understanding of galaxy formation and evolution within the Universe \citep[e.g.,][]{Alex+12, Krawczynski+13, Combes2021}.

There is increasing evidence that black holes are prevalent in large galaxies in the local Universe \citep[e.g.,][]{Kormendy+13}. AGN display a wide range of luminosities and other properties, with an expanding body of observational data \citep[e.g.,][]{Rosas+16, Lyu+24} providing essential constraints for physical models. However, a relatively unexplored area involves moderate-luminosity AGN, which occupies an intermediate range between low-luminosity \citep[e.g.,][]{Jeon+23} and high-luminosity AGN. The study of growing black holes and their host galaxies, especially those with moderate luminosities (\( 10^{44} \leq L_\mathrm{BOL}\, \mathrm{[erg\, s^{-1}]} \geq 10^{45} \)), offers critical insights into the co-evolution of galaxies and their central supermassive black holes \citep{Simmons+11}.

Statistical analysis of the clustering of AGN aids in suggesting the particular underlying processes responsible for their different observational correlations, which in turn offers insights into their formation and their nuclear activity \citep[e.g.,][]{Oogi+20,Chaves+16}. The AGN-halo mass connection could be studied by clustering measurements to investigate the presence of a predominant triggering mechanism. in this case. At least in the case of high X-ray luminosity AGN, it is rather well established that they result from galaxy mergers \citep[e.g.][]{Springel+05, Lin+23}, with more massive galaxies exhibiting greater nuclear activity. Using the method of halo occupation distribution (HOD) it is possible to understand whether the host galaxies have only one phase of activation or experience multiple episodes, and of what type \citep[e.g.,][]{Krumpe+23}.

Computer simulations are essential in advancing our knowledge of the formation and evolution of galactic structures and their cosmological context \citep{Zana+19,Wein+18,Schaye+23}, as they provide a crucial reference for corroborating observations \citep{Angulo+22} and predictions of future galaxy surveys \citep{Hadzhiyska+20}. Hydrodynamic simulations allow us to study behavior of dark matter \citep[and references therein]{Yuan+22} and, in particular, the properties of specific populations of supermassive black holes (SMBHs) and AGN \citep[e.g.,][]{Musoke+20,Li+23}. Measuring the space density of AGN across cosmological volumes provides valuable insights into the growth history of SMBHs in the local universe \citep{Laloux+23,Nandra+05}, and comparing them with the data provided by simulations can further our understanding of such process.

Moderate X-ray luminosity AGN represent a stage in the lifecycle of black hole growth that lies between low-luminosity \citep[e.g.][]{Hopkins+06}, quiescent AGN and high-luminosity, and rapidly accreting AGN. By focusing on mXAGN, we gain insights into this transitional phase, helping to piece together a more complete picture of how black holes grow over cosmic time \citep[e.g.][]{Kocevski+12} within the hierarchical framework of galaxy formation. Moderate X-ray luminosity AGN are found in a variety of cosmic environments \citep[e.g.][]{Kauffmann+03,Rey+21}, from isolated field galaxies to galaxy groups and clusters \cite[e.g.][]{Alle+14}. Their presence in different environments allows us to study how large-scale structures influence AGN activity. Since galaxy mergers, interactions, and dense environments can trigger AGN \cite[e.g.][]{Allevato+11}, examining these galaxies provides a window into the environmental factors that fuel or suppress AGN activity, which are critical for understanding hierarchical growth on different scales. By observing their distribution and specific BH accretion rates \cite[e.g.][]{Hirschmann+14,Aird+18, Schulze+18}, AGN can be used as tracers of large-scale structures and offer insight into the relationship between AGN activity and the underlying dark matter in such structures \citep[e.g.][]{Leauthaud+15}.

In this work, we use public data from the Evolution and Assembly of GaLaxies and their Environments (\eagle{}\footnote{\url{https://icc.dur.ac.uk/Eagle/}}) simulation \citep{Crain+15,Schaye+15} to investigate the clustering of mXAGN using the HOD method, and compare our results with available observations. The content of the paper is as follows: Section \ref{sec:methodology} describes the selection of our sample of mXAGN from the \eagle{} simulation, as well as providing basic information about it. Our main results are discussed in Section \ref{sec:results}. Finally, closing remarks and conclusions are provided in Section \ref{sec:conclusions}. 

%%%%%%%%%%%%%%%%%%%%%%%%%%%%%%%%%%%%%%%%%%%%%%%%%%%%%%
\section{Methodology}\label{sec:methodology}
%%%%%%%%%%%%%%%%%%%%%%%%%%%%%%%%%%%%%%%%%%%%%%%%%%%%%%

This section discusses the fundamental properties of the \eagle{} simulation and outlines the process used to identify mXAGN. Additionally, we examine the merger history of mXAGN to identify past mergers and provide details about the criteria for constructing the HODs, considering both satellite and central galaxies.

%%%%%%%%%%%%%%%%%%%%%%%%%%%%%%%%%%%%%%%%%%%%%%%%%%%%%%
\subsection{The \eagle{} simulation}\label{sec:eaglesim}
%%%%%%%%%%%%%%%%%%%%%%%%%%%%%%%%%%%%%%%%%%%%%%%%%%%%%%

\eagle{} provides a powerful framework for studying mXAGN in the context of hierarchical structure formation and large cosmic structures \citep[e.g.,][]{McAlpine+20,Saha+24}. Due to its size, (max length size $L=100$ cMpc, comoving Mpc), mass ($m \sim 10^6$\,M$_\odot$) and spatial ($\approx 2$ ckpc) resolution, and the wide array of physical processes taken into account to build the simulation, \eagle{} allows for tracking black hole accretion and AGN activity over time, providing valuable insights into how mXAGN evolve within the broader cosmic framework.

The \eagle{} simulation suit used a modified N-Body Tree-PM smoothed particle hydrodynamics (SPH) code GADGET\footnote{\url{https://wwwmpa.mpa-garching.mpg.de/gadget/}} \citep{Springel+01,Springel+05,Springel+21}.  \eagle{} adopts the following values for the standard cosmological parameters: $\Omega_{\rm m}= 0.307$, $\Omega_{\Lambda} = 0.693$, $\Omega_{\rm b} = 0.04825$, ${\rm h} = 0.6777$, $\sigma_{8} = 0.8288$, and ${n}_{\rm s} = 0.9611$ \citep{Planck+14}. \eagle{}  incorporates relevant physical processes such as radiative cooling and photoionization heating \citep{Wiersma+09a}, star formation \citep{Schaye+08}, stellar mass loss \citep{Wiersma+09b}, stellar feedback \citep{Dalla-Vecchia+12}, and BH growth and feedback \cite[e.g.,][]{Springel+05,Booth+10, Wang+15,Byrne+23}. 
      
To carry out our analysis, we use data specifically from the \texttt{Ref-L100N1504} run. This simulation has a box size of 100 cMpc and 1504$^{3}$ particle number. The Structured Query Language (SQL) interface \citep{McAlpine+16} was used to extract the properties of galaxies and groups within the \eagle{} simulations.

%%%%%%%%%%%%%%%%%%%%%%%%%%%%%%%%%%%%%%%%%%%%%%%%%%%%%%
\subsection{Identifying AGN} \label{sec:xagn}
%%%%%%%%%%%%%%%%%%%%%%%%%%%%%%%%%%%%%%%%%%%%%%%%%%%%%%
Using the simulation data, selection criteria are applied to identify mXAGN that have undergone merger events, activating these galaxies. Subsequently, their merger history is analyzed to classify them as minor or major. The first step of the selection criteria is to apply a cut in X-ray luminosity to select only galaxies that emit moderate X-ray luminosity, $L_{\rm X}\approx10^{43}\,\mathrm{erg\,s^{-1}}$.
Following \cite{McAlpine+17}, we connect the BH accretion rate (BHAR), $\dot{m}$, provided by the data simulation with the bolometric luminosity using the equation

\begin{equation}
\label{eq:l_bol}
L_\mathrm{bol}=\epsilon \dot{m} {\rm c}^{2} \,
\end{equation}

where $L_\mathrm{bol}$ is the bolometric luminosity and radiative efficiency assumed of $\epsilon$= 0.1 \citep{Crain+15,Dave+19, Bennett+24}.

To estimate the soft X-ray luminosity (0.5--2 keV), we substitute the bolometric luminosity obtained from Equation \ref{eq:l_bol} into the relation proposed by \cite{Lusso+12}:

\begin{equation}
\label{eq:lusso}
\log[L_\mathrm{bol}/L_\mathrm{band}] = a_{1} x+ a_{2} x^2 + a_{3} x^3 + b,
\end{equation}

where $L_\mathrm{band}$ correspond to the 0.5--2 keV band luminosity, x = $\log$ $L_\mathrm{bol}$$-$12, $a_\mathrm{1}$=0.248, $a_\mathrm{2}$=0.061, $a_\mathrm{3}$=-0.041 and $b$=1.431 are bolometric correction coefficients.

%%%%%%%%%%%%%%%%%%%%%%%%%%%%%%%%%%%%%%%%%%%%%%%%%%%%%%
\subsection{Merger Event} \label{sec:merger}
%%%%%%%%%%%%%%%%%%%%%%%%%%%%%%%%%%%%%%%%%%%%%%%%%%%%%%
The following selection criteria were employed to quantify the number of galaxies that experienced either a minor or major merger as a triggering mechanism. 
Using the merger tree from the \eagle{} simulation, we traced the last progenitor \texttt{LastProgID} parameter of the galaxies identified as mXAGN (based on the previously mentioned criteria). Additionally, we examined the merger history of the progenitor galaxies to determine whether they had experienced any mergers. To carry out this analysis, the following parameters need to be defined:

\begin{enumerate}
    \item Parent 1 or main progenitor (P$_{1}$): \texttt{GalaxyID} of the last subhalo progenitor of the galaxy selected as active as well as the \texttt{GroupID} to which main parent galaxy belongs (parent group).
    
    \item Group parent (GP): \texttt{GroupID} of the group parent is defined as the group in which the main progenitor is a member. 
    
    \item Parent 2 or secondary progenitor (P$_{2}$): \texttt{GalaxyID} closest member of the group parent to the main subhalo progenitor.
\end{enumerate}

First, we calculated the distance, $d_{p}$, from each parent group member to the main parent following the relation

\begin{equation}
    \label{eq:dp_half_mass}
     d_{p} \leq 5\,R_{M_*}
\end{equation} 

where $R_{M_*}$ is the physical radius enclosing half of the stellar mass given by \eagle{} simulation \citep{Qu+17}. 

Secondly, we use the distance obtained from Equation \ref{eq:dp_half_mass} as an indicator to assess whether the progenitor pair may have undergone a potential merger. We then classify mergers as major or minor at the redshift where both progenitors are observed. We define the mass ratio of a galaxy merger as \( \mu = M_\mathrm{2}/M_\mathrm{1} \), where \( M_\mathrm{1} > M_\mathrm{2} \). Mergers with mass ratios in the range \( 0.25 \leq \mu \leq 1.0 \) are classified as major mergers, while those with \( 0.1 \leq \mu < 0.25 \) are classified as minor mergers. Additionally, the criteria outlined by Faruki \& Shapiro \citep[F-S;][]{Farouki+81} must be satisfied to identify a merger event between two subhalo progenitors:

\begin{enumerate}
    \item Their relative velocity \(V_\mathrm{12} = \vert V_\mathrm{1} - V_\mathrm{2} \vert \) is less than average of velocity dispersion {\sc rms} of both subhalos $\langle V_\mathrm{rms} \rangle$; i.e., $V_\mathrm{12}\leq \langle V_\mathrm{rms} \rangle$. 

    \item Their relative physical separation \(R_\mathrm{12}=\vert r_\mathrm{1}- r_\mathrm{2}\vert \) is less than the sum of the virial radius of both subhalos: $R_\mathrm{12} \leq R_\mathrm{v1} + R_\mathrm{v2}$.
    
\end{enumerate}

Satisfying the F-S criteria strongly indicates that a merger is likely to occur, typically within one giga-year after the criteria are met. Although the F-S criteria were derived from dynamical simulations, we use them here as a proxy for strong interactions or mergers in the cosmological context. We also verified that all subhalos meeting these criteria eventually merged.

%%%%%%%%%%%%%%%%%%%%%%%%%%%%%%%%%%%%%%%%%%%%%%%%%%%%%%
\subsection{HOD} \label{sec:hod}
%%%%%%%%%%%%%%%%%%%%%%%%%%%%%%%%%%%%%%%%%%%%%%%%%%%%%%
To model our HOD, we use the \texttt{GroupMass} parameter provided by the \eagle{} simulation, which represents the total mass of the halo (M$_{G}$) as defined by the Friends-of-Friends (FoF) algorithm. We select all halos with \(M_{G} \geq 10^{12.75}\,\mathrm{M}_{\odot} \, h^{-1}\) and then extract two distinct galaxy catalogs from these selected halos at 10 different redshifts: 0, 0.18, 0.27, 0.5, 0.62, 1.0, 1.24, 1.48, 1.74, and 2.0. One catalog contains only satellite galaxies (SGC), while the other includes only central galaxies (CGC) as identified in our analysis. This classification is based on the \texttt{SubGroupNumber} parameter extracted directly from the simulation data, where \texttt{SubGroupNumber} = 0 identifies central galaxies, and \texttt{SubGroupNumber} $>$ 0 identifies satellites. We restricted galaxies that have black hole mass \(M_{\rm BH}\,>\,10^{6}\,\mathrm{M}_{\odot} h^{-1}\), and stellar mass \(M_{\star}\,>\,10^{8}\,\mathrm{M}_{\odot} h^{-1}\) represents the stellar content of a subhalo within a 30 kpc spherical aperture \citep{Schaye+15}.

As previously defined in \cite{Altamirano+16}, the HOD has been implemented as the number of occupations in each group size halo that contain a galaxy with an AGN considering the relation

\begin{equation}
	\label{eq:hod_lily}
	N(M_\mathrm{G}) = \frac{n_\mathrm{G_\mathrm{AGN}}} {n_\mathrm{G}},
	\end{equation}

where \(n_{G}\) represents the number density of groups halo mass \textbf{\(10^{12.75}\,\leq\,M_{G}\,[\mathrm{M}_{\odot} h^{-1}]\,\geq\, 10^{14.35}\)} in the \eagle{} simulation and \(n_\mathrm{G_\mathrm{AGN}}\) is the number density of groups with halo mass \textbf{\(10^{12.75}\,\leq\,M_{G}\,[\mathrm{M}_{\odot} h^{-1}]\,\geq\, 10^{14.35}\)}with at least one mXAGN.

%%%%%%%%%%%%%%%%%%%%%%%%%%%%%%%%%%%%%%%%%%%%%%%%%%%%%%
\section{Results and Discussion}\label{sec:results}
%%%%%%%%%%%%%%%%%%%%%%%%%%%%%%%%%%%%%%%%%%%%%%%%%%%%%%

We examined the HOD shape based on the \eagle{} simulation \texttt{RefL0100N1504} and found it very similar to the one described in \cite{Altamirano+16}. The primary distinction is that our earlier analysis was exclusively based on dark matter halos. In contrast, this new study integrates cosmological hydrodynamical simulations, in which dark and baryonic matter play critical roles.

%!********************************************
\begin{figure}
\centering
      \includegraphics[width=1.0\linewidth]{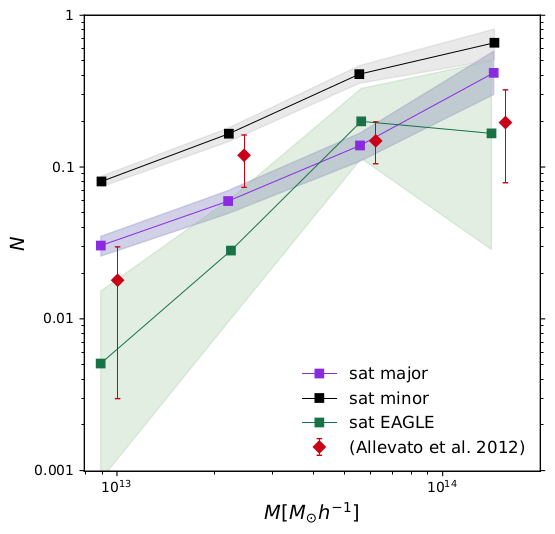}
      \caption{HOD for satellites at $z=0.5$. The results derived from the \eagle{} simulation are displayed in green. In contrast, the findings from the GADGET-2 simulations \cite{Altamirano+16}, corresponding to major and minor mergers, are presented in blue and black, respectively. The observational results at $\Bar{z}$=0.5 from \cite{Allevato+12} are shown in red (See online manuscript for color graph).}
      \label{fig:occ} 
\end{figure}
%!********************************************

In Fig. \ref{fig:occ}, we show the result of our analysis using the \eagle{} \texttt{RefL0100N1504} simulation, considering only satellites (in green). In this figure, we compare the contribution of the satellites in the halo occupation function at \(z=0.5\), such satellites were extracted as described in Section \ref{sec:methodology}. The errors in Fig. \ref{fig:occ} were derived using the Equations 7 \& 12 of \cite{Ge+86}. 

%%%%%%%%%%%%%%%%%%%%%%%%%%%%
\begin{figure*}
\centering
     \includegraphics[width=0.85\linewidth]{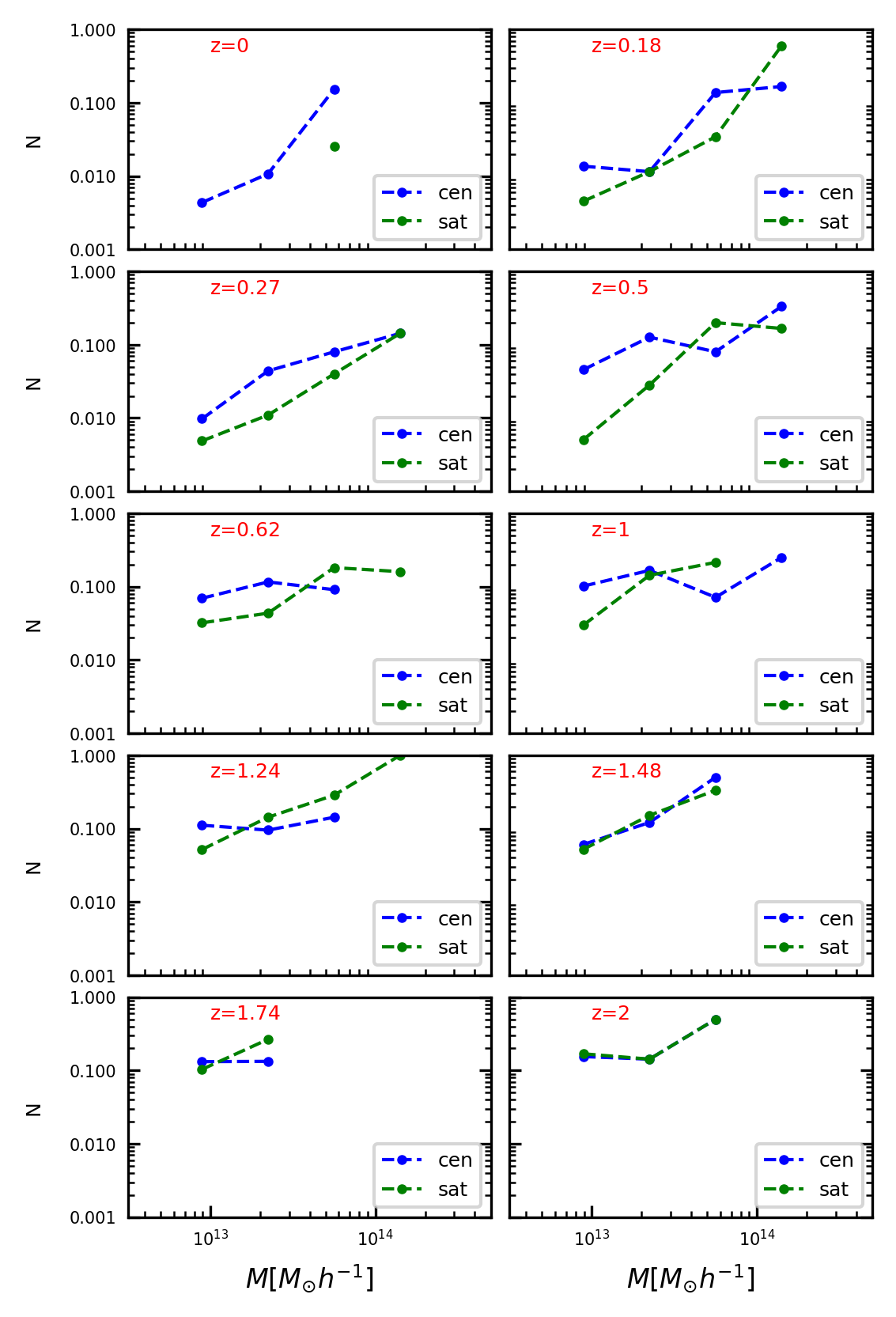}
    \caption{The HOD of mXAGN at redshift between 0 and 2. is illustrated for both central and satellite galaxies. In the plot, satellite galaxies are represented in blue, while central galaxies are depicted in green. These galaxies are located within halos with typical masses associated with groups and clusters.}
\label{fig:multiplot} 
\end{figure*}
%%%%%%%%%%%%%%%%%%%%%%%%%%%%%

In Fig. \ref{fig:multiplot}, we present the HOD calculated at various redshifts. The central galaxies hosting an mXAGN are plotted in green, while the satellite galaxies hosting an mXAGN, as previously defined in Section \ref{sec:xagn}, are represented in blue. It is significant that the number density of subhalos, including both central and satellite galaxies, is low at \(z=0\). This result is consistent with the predictions made by \cite{Mishra+20} where AGN are lower in large structures like clusters. Despite the increased likelihood of mergers at this redshift, the efficiency of detecting AGN within these mass group-like halos is lower than that at higher redshifts.

We plotted the HOD profiles across different redshifts (Fig. ~\ref{fig:multiplot}), applying a consistent methodology to the satellite (SGC) and central galaxies (CGC), constructed as described in Section \ref{sec:hod}. It is important to note that at higher redshifts, the number density of AGN with moderate luminosity is significantly greater compared to lower redshifts. This substantial increase in AGN density during earlier cosmic epochs suggests that these moderate luminosity AGN were much more prevalent in the early stages of galaxy formation and evolution, potentially indicating a period of heightened accretion activity onto supermassive black holes.

Table~\ref{table:1} presents the numerical densities detected at four key redshifts: 0, 0.5, 1 and 2. These densities represent the frequency of occurrences of satellite and central galaxies within our dataset at each redshift. At \(z=0\), which corresponds to the present day in cosmological terms, our analysis indicates that recent mergers are not significantly influencing the galaxies in our sample. This suggests that earlier mergers may have been more effective in triggering these galaxies, consistent with the findings of \cite{Ellison+19}.

At the intermediate \(z=0.5\), roughly corresponding to a look-back time of about 5 billion years, we observed that approximately \(56\%\) of the galaxies in our sample had undergone minor mergers for satellites while \(50\%\) for centrals. This merger fraction at redshift 0.5 highlights the importance of minor mergers in molding galaxies during this period. The increased merger activity could be attributed to the more crowded and dynamic environment in the earlier Universe, leading to more frequent gravitational interactions between galaxies, and also to the amount of available gas to ignite the SMBHs.

At \(z=1\), representing a look-back time of about 8 billion years, the contribution of the fraction of major mergers as a triggering mechanism shows an increase to approximately 50\% for satellites and 62\% for central mXAGN. At this redshift, the number of central mXAGN substantially increases compared to the number of satellite mXAGN identified. This notable rise indicates that central galaxies are experiencing a more pronounced phase of activity at this stage, suggesting that their environments or merger histories may play a crucial role in fostering AGN activity compared to their satellite counterparts. 

At \(z=2\), while the fraction of mXAGN is on the rise, minor mergers are the predominant type of interaction occurring among these galaxies. This prevalence of minor mergers suggests that, although more galaxies are becoming active during this epoch, the interactions primarily involve smaller systems merging with larger ones rather than significant major mergers. Such minor mergers may play a crucial role in the gradual evolution of these galaxies, potentially contributing to their growth and the triggering of AGN without the dramatic changes associated with major merger events. This dynamic interplay indicates a complex evolutionary landscape in which galaxy activity is influenced by the nature of their interactions.

\begin{table}[!t]\centering
\small
  	\caption{Number densities for satellite and central mXAGN.}
 	\begin{tabular}{ ccccc }
 	\hline
        Redshift & $n_\mathrm{AGN(sat)}$ & $n_\mathrm{AGN(cent)}$ \\ [0.5ex] 
    \hline
	    0   & 1$\times$ 10 $^{-6}$ &8$\times$ 10 $^{-6}$ \\
	    0.5 & 9$\times$ 10 $^{-6}$  &2.2$\times$ 10 $^{-5}$ \\
	      1.0 & 1.4$\times$ 10 $^{-5}$ &2.6$\times$ 10 $^{-5}$\\ 
            2.0 & 1.4$\times$ 10 $^{-5}$ &1.3$\times$ 10 $^{-5}$\\
       %[1ex] 
	\hline          
    \end{tabular}
     \label{table:1}
\end{table}

As shown in Fig.~\ref{fig:occ}, the data points fell within the central and satellite AGN error bars, indicating no significant difference between the two samples. We performed a Kolmogorov-Smirnov test to evaluate whether the two samples were statistically distinct. However, the test did not yield substantial results, likely due to the limited sample sizes, which may have diminished the statistical power and the ability to detect significant differences.

%!********************************************
\begin{figure}
\centering
      \includegraphics[width=1.0\linewidth]{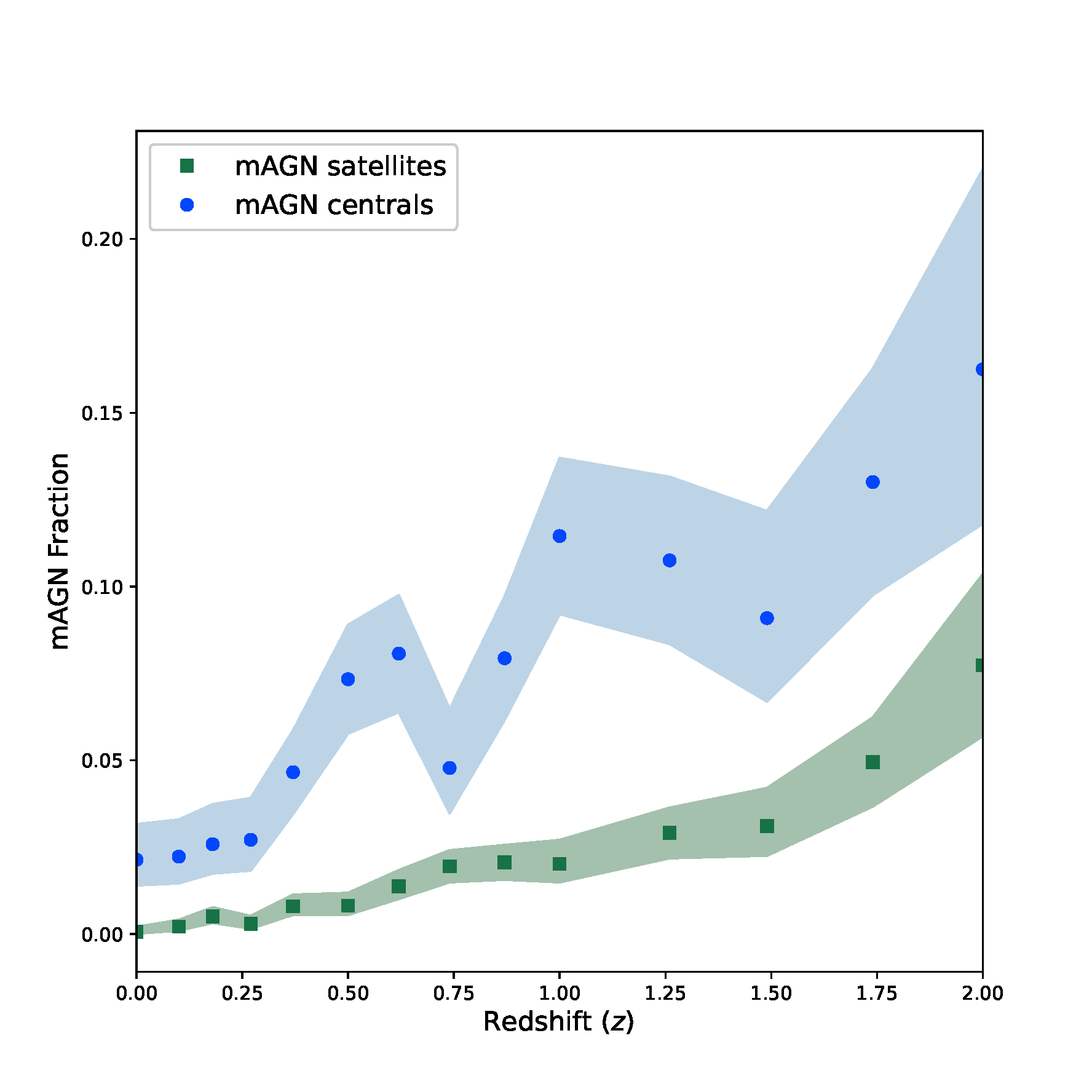}
      \caption{Fraction of mXAGN as a function of redshift hosted by halo mass \(M_{G}\,\geq\,10^{12.75}\,\mathrm{M}_{\odot} h^{-1}\), based on detections obtained from the analyzed simulation.} In blue color points are central mXAGN and in green color points are satellite mXAGN. The shaded area represents the 1$\sigma$ error.
      \label{fig:fract} 
\end{figure}
%!********************************************

In Fig.~\ref{fig:fract} we present the fraction of mXAGN in the typical halo mass of groups with \(M_{G}\,\geq \,10^{12.75}\,\mathrm{M}_{\odot} h^{-1}\) across a redshift range of \(0<z<2\). Our analysis reveals that the fraction of central and satellite active galaxies tends to increase at higher redshifts. This trend suggests that the activity of these types of galaxies is more pronounced in the earlier stages of the universe, indicating a potentially significant evolutionary phase for both central and satellite galaxies during this period. 

\begin{figure}
\centering
     \includegraphics[width=1.0\linewidth]{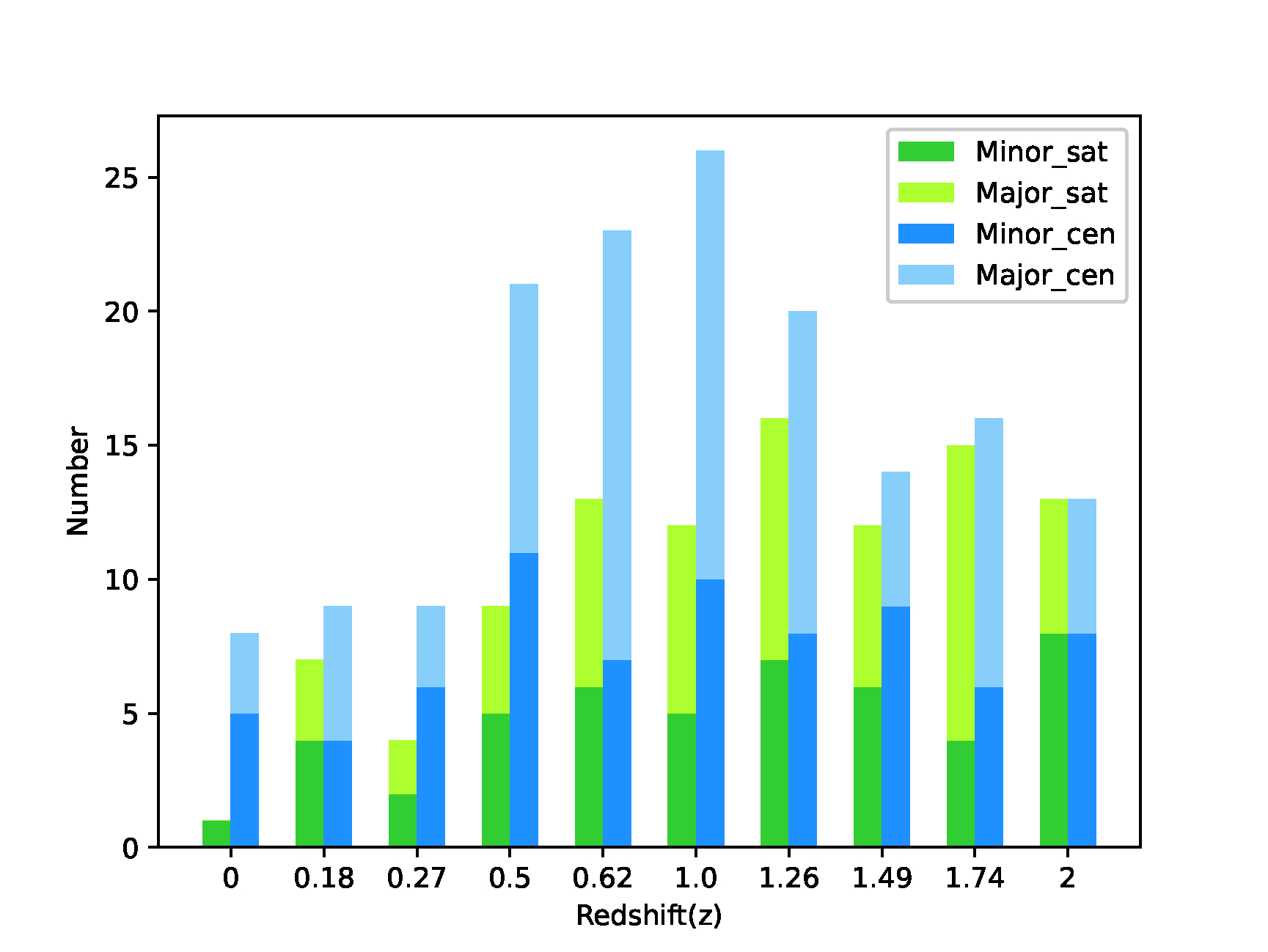}
    \caption{Histogram showing the number of mXAGN present in our catalogs of central and satellite galaxies, resulting in major and minor mergers, respectively, evaluated at 10 different redshifts. In each case, the left column corresponds to the mXAGN from the SGC, and the right column corresponds to the CGC. In each column, a light color is used to indicate those mXAGN resulting from minor mergers, while a solid color is used to indicate major mergers.}
\label{fig:histogram} 
\end{figure}

A comparison between the minor and major contributions in each redshift for all the mXAGN in our catalogs of central and satellite galaxies is shown in Fig.~\ref{fig:histogram}. It is found that minor mergers are the primary mass contributors in central massive mXAGNs  (\(M_{*}/M_\odot \approx 10^{11}-10^{12}\)) at \(z=0\). At \(z=1\), it is found that both minor mergers and minor mergers in mAGNs (\(M_{*}/M_\odot \approx 10^{10}-10^{12}\)) contribute to the total stellar mass. Finally, for \(z=2\), t is found that minor mergers are the main contributors to the stellar mass increase of mAGNs in the range of (\(M_{*}/M_\odot \approx 10^{10}-10^{11}\)), in concordance with \cite{Qu+17} at \(z=1\) and \(z=2\), but not for \(z=0\) in each bins of massive central galaxies using \eagle{} simulation.

%!********************************************
%!********************************************

%%%%%%%%%%%%%%%%%%%%%%%%%%%%%%%%%%%%%%%%%%%%%%%%%%%%%
\section{Conclusions}\label{sec:conclusions}
%%%%%%%%%%%%%%%%%%%%%%%%%%%%%%%%%%%%%%%%%%%%%%%%%%%%%%

We used the public cosmological hydrodynamical simulation \eagle{} to construct the HOD for mXAGN (0.2--5 keV) AGN within galaxy groups with masses greater than \( 10^{12.75} \, h^{-1} {\rm M}_{\odot}\) at different redshifts, with particular emphasis on \(z=0.5\) to compare with previous results and observations to this type of AGN. 

We compared our current results with those of \cite{Altamirano+16} to assess the impact of baryonic components in the sub-grid models included in the \eagle{} simulation, which covers a large volume of 100 cMpc$^{3}$, on the measurements of clustering in group/cluster-like mass halos. Our current findings are consistent in general with our previous results. At redshift 0.5, the shape of the HOD for satellite galaxies hosting moderate-luminosity AGN tends to flatten at higher halo masses.

The \eagle{} simulation reveals a significant decline in the number of high-mass halos, indicating a reduction in the presence of galaxies that host mXAGN at these higher masses. Consequently, our analysis is based on a number-limited sample of mXAGN, which constrains our ability to draw statistical conclusions about their properties and distribution. This limitation highlights the need for more extensive simulation data to better understand the behaviour of mXAGN in high-mass halos. 

The data presented in Table~\ref{table:1} highlight the evolving role of mergers in galaxy evolution across different cosmic epochs, in particular here for mXAGN. At \(z=0\), no recent mergers are shown, while at redshifts \(z=0.5\) and \(z=2\), the minor merger appears to be the predominant mechanism. While at \(z=1\), the major merger is crucial to ignite mXAGN. This underscores the ongoing importance of mergers in triggering mXAGN throughout the Universe's history.

We calculated the HOD for satellites and central galaxies that host mXAGN for a given range of redshifts, obtaining lower number densities at \(z=0\) than a high redshift \(z=2\). We studied the merger history of our sample of mXAGN at the redshift range and found that minor mergers play a crucial role in almost all redshifts. This finding underscores the importance of these smaller interactions in shaping galaxy evolution and activity throughout cosmic time. Our analysis suggests that even as the Universe transitions through different epochs, minor mergers consistently contribute to the growth and dynamical evolution of galaxies.

In future work, it will be essential to compare our numerical results with extensive observational data to evaluate the reliability of the models incorporated into the simulations. By contrasting our simulations with large volumes of observational data, we can assess the accuracy and robustness of the theoretical models employed. This comparison will not only validate the models but also enhance their predictive power, leading to more accurate forecasts of observational outcomes. Such rigorous testing is crucial for refining our understanding of the underlying physical processes and ensuring that the simulations provide meaningful insights into real-world phenomena.

Major contributions are expected to be made by \flamingo{}\footnote{
\url{https://flamingo.strw.leidenuniv.nl/}} \citep{Schaye+23}, a state-of-the-art cosmological simulation. \flamingo{} encompasses a much larger simulation volume and higher resolution in terms of particle count compared to \eagle{} and incorporates the latest physical models and cosmological parameters. Thus, \flamingo{} will soon enable researchers to explore the dynamics of gas, galaxy evolution, and the influence of mXAGN on cosmic structure formation. By integrating dark matter, baryonic matter, and neutrinos, \flamingo{} will provide a framework for understanding complex interactions shaping large-scale cosmic structures and their growth across time.

%%%%%%%%%%%%%%%%%%%%%%%%%%%%%%%%%%%%%%%%%%%%%%%%%%%%%%
\section*{Acknowledgments}
%%%%%%%%%%%%%%%%%%%%%%%%%%%%%%%%%%%%%%%%%%%%%%%%%%%%%%

This research was funded by a CONAHCyT postdoctoral fellowship, CONAHCyT Ciencia B\'asica grant 252531, and PAPIIT IN114423. LAD acknowledges the Institute of Astronomy for providing support and a workspace during her postdoctoral research and thanks Mauricio Elias for his valuable comments. We thank the Virgo Consortium for making their simulation data available. The \eagle{} simulations were performed using the DiRAC-2 facility in Durham, managed by the ICC, and the PRACE facility Curie, based at TGCC, CEA, Bruyères-le-Ch\^{a}tel, France. The public version of ChatGPT \citep{ChatGPT2023} was used to review the basic spelling and grammar of the document.

%%%%%%%%%%%%%%%%%%%%%%%%%%%%%%%%%%%%%%%%%%%%%%%%%%%%%%%

\bibliographystyle{elsarticle-harv}
%\bibliography{biblio}

\end{document}